\documentclass[final,5p,times,twocolumn]{elsarticle}

\usepackage{amssymb}

\usepackage{todonotes}

\usepackage{xspace}
\usepackage{orcidlink}

\usepackage{amsmath}
\usepackage{multirow}

\usepackage{algorithm}
\usepackage{algpseudocode}

\usepackage{booktabs}

\newcounter{bla}

\journal{Computer Physics Communications}

\newcommand{\pvfl}{{Popcorn Void Finder}\xspace}
\newcommand{\pvf}{{\textit{PVF}}\xspace}

\newcommand{\vsfl}{{void-size function}\xspace}
\newcommand{\vsf}{{\textit{VSF}}\xspace}

\newcommand{\vgcfl}{{void-galaxy correlation function}\xspace}
\newcommand{\vgcf}{{\textit{VGCF}}\xspace}

\newcommand{\nnnl}{{Nearest-Neighbour Search}\xspace}
\newcommand{\nnn}{{\textit{NNS}}\xspace}

\newcommand{\cxx}{{\texttt{C++}}\xspace}
\newcommand{\serafin}{{\texttt{serafin}}\xspace}

\definecolor{jbc}{RGB}{69, 122, 17}
\definecolor{anr}{RGB}{255, 127, 0}

\newcommand{\del}[1]{{\iffalse {#1} \fi}}

\begin{document}

\begin{frontmatter}



\title{Refined, Reliable and Reproducible: Algorithmic Analysis and Reimplementation of the Popcorn Void Finder}


\author[iate,famaf]{Schachner, Alvaro Roy \orcidlink{0009-0008-5460-8720}\corref{alvaro}}
\author[iate,oac]{Paz, Dante Javier \orcidlink{0000-0001-6094-663X}}
\author[computaex]{Rico Gallego, Juan Antonio \orcidlink{0000-0002-4264-7473}}
\author[mpe]{Correa, Carlos Mauricio \orcidlink{0000-0003-1596-5254}}
\author[iate,oac]{Ruiz, Andrés Nicolás \orcidlink{0000-0001-5035-4913}}
\author[iate,conicet]{Gualpa, Sebastián Rogelio \orcidlink{0009-0008-4021-8561}}
\author[conae,conicet,famaf]{Cabral, Juan Bautista \orcidlink{0000-0002-7351-0680}}

\cortext[alvaro] {Corresponding author.\\\textit{E-mail address:} alvaro.schachner@unc.edu.ar}

\address[iate]{Instituto de Astronomía Teórica y Experimental, CONICET-UNC, Laprida 854, X5000BGR Córdoba, Argentina}

\address[famaf]{Facultad de Matemática, Astronomía, Física y Computación, Universidad Nacional de Córdoba, Av. Medina Allende s/n, X5000HUA Córdoba, Argentina}

\address[oac]{Observatorio Astronómico de Córdoba, Universidad Nacional de Córdoba, Laprida 854, X5000BGR Córdoba, Argentina}

\address[computaex]{COMPUTAEX, Extremadura Supercomputing Center, Ctra Nacional 521, km 41.8, 10071 Cáceres, España}

\address[mpe]{Max-Planck-Institut für extraterrestrische Physik, Postfach 1312, Giessenbachstr, D-85748 Garching, Germany}

\address[conicet]{Consejo Nacional de Investigaciones Científicas y Técnicas (CONICET), Godoy Cruz 2290, C1425FQB Ciudad Autónoma de Buenos Aires, Argentina}

\address[conae]{Gerencia de Vinculación Tecnológica, Comisión Nacional de Actividades Espaciales (CONAE), Falda del Cañete, Córdoba, Argentina}

\begin{abstract}

Cosmic voids are vast underdense regions of the Universe that encode valuable information about its large-scale structure and evolution.
By extending spherical voids into overlapping sphere structures, the Popcorn Void Finder demonstrated an improved statistical characterization of cosmic void populations.
However, its origin as a research prototype progressively hindered its maintainability and extensibility.
Moreover, post-publication algorithmic modifications have not been systematically evaluated, leaving their statistical and computational consequences insufficiently characterized.
In this work, we present a comprehensive reimplementation of the Popcorn Void Finder, prioritizing algorithmic transparency, maintainability, and reproducibility.
We document the complete identification procedure, revisit the original algorithmic choices, and evaluate their statistical and computational consequences.
Comparisons between implementations revealed a significant change in the behavior of the Void Size Function, which is now characterized by a power-law regime transitioning to a Schechter-like distribution.
These changes not only provide a better characterization of void sizes, but enable a new and more robust method to the determination of the shot-noise dominated regime, without implicit dependencies on internal finder parameters present in previous works.
Performance benchmarks demonstrate a $1.62\times$ speedup in spherical void identification and a substantial improvement in overlap-removal throughput.
We determined the Popcorn algorithm to be negatively affected by load imbalance, and found a speedup of $63\times$ in a tested configuration with a newly introduced early growth termination criterion.
These results establish a more maintainable and reproducible foundation for popcorn void identification, while clarifying the statistical and computational trade-offs that govern its performance.
\\


\noindent \textbf{PROGRAM SUMMARY}

\begin{small}
\noindent
{\em Program Title:} \pvfl \\
{\em CPC Library link to program files:} (to be added by Technical Editor) \\
{\em Developer's repository link:} \url{https://gitlab.com/void-research-labs/popcorn-void-finder} \\
{\em Licensing provisions:} BSD 3-clause \\
{\em Programming language:} \cxx \\
{\em External routines/libraries:} OpenMP, MPI, HDF5 (optional) \\
{\em Nature of problem:}
Cosmic void boundaries are not uniquely defined by observable quantities, and their identification therefore requires an explicit model.
Spherical void finders provide a widely used approach to characterizing void populations statistically, but their geometric assumptions impose important limitations.
In particular, complex underdense regions may be fragmented into multiple spherical voids, especially when their morphology is constrained by surrounding high-density structures.
This fragmentation can distort the inferred size distribution and abundance of voids, while discarding information about the geometry and connectivity of the underlying underdense regions.
These limitations motivate the development of void-identification methods that retain the computational and statistical advantages of spherical approaches while providing a more flexible representation of complex void morphologies. \\
{\em Solution method:}
The Popcorn Void Finder generalizes spherical voids by representing each structure as the union of an arbitrary number of overlapping spheres, allowing complex and extended underdense regions to be described without imposing a fixed global geometry. The identification process begins with a catalogue of spherical voids, which serve as initial seeds for popcorn structures. Each seed is iteratively expanded by identifying and adding spheres that satisfy the void underdensity criterion while maximizing the additional volume contributed to the structure. This process builds progressively more complex configurations until no further suitable spheres can be incorporated. The resulting structures retain the statistical foundations of spherical void identification while providing a flexible representation of non-spherical void morphologies.
%

\begin{keyword}
Cosmic voids \sep Large-scale structure of Universe \sep Void identification algorithms \sep
Computational cosmology \sep High-performance computing \sep Scientific software
\end{keyword}

\end{small}
   \end{abstract}
\end{frontmatter}


\section{Introduction}
\label{sec:intro}

Cosmic voids are vast underdense regions in space, appearing as large, nearly empty spaces between filaments of galaxies, making up the biggest observable structures of the Universe \citep[e.g. ][ and references there in]{2004.Seth.Hierarchy, 2026Contarini}.
Their statistical and geometrical properties have been used to constrain cosmological parameters, investigate the growth of structure, test alternative theories of gravity, and study the nature of dark energy \citep{cai+2015,2019Correa,aubert+2022,hamaus+2022,2022.Contarini.VSF,2023.Radinovich.VGCF}.
Furthermore, void environments are less affected by several baryonic processes \citep{2022Rodriguez} that complicate the interpretation of dense regions, making them especially attractive for precision cosmology analyses \citep{2023.Correa.Voids,zvoidsI,zvoidsII}.

Several statistical observables derived from cosmic void populations have proven to be valuable tools for modern cosmology.
Among the most relevant examples are the \vsfl (\vsf) and the \vgcfl (\vgcf).
The \vsf describes the abundance of voids as a function of their effective radius, in close analogy with the mass function used for dark matter halos.
Its shape and evolution are sensitive to the underlying cosmological model, the amplitude of matter fluctuations, and the growth history of large-scale structure \citep{2004.Seth.Hierarchy,zvoidsI,2013Jennings}.
Similarly, the \vgcf characterizes the statistical distribution of galaxies around void centers, providing information about the density and velocity profiles surrounding underdense regions.
This observable has become particularly relevant in observational cosmology due to its sensitivity to redshift-space distortions and large-scale velocity flows \citep{2013.Paz.Cluesii,hamaus+2015,Massara2022,zvoidsII}.
Through the Alcock-Paczynski test and redshift-space distortion measurements, the \vsf and \vgcf have been used to constrain the expansion history of the Universe and the growth rate of cosmic structure as well as to test alternative gravity models and dark energy scenarios \citep{2013.Paz.Cluesii,cai+2015,aubert+2022,hamaus+2022,2023.Radinovich.VGCF}.

Together, these applications highlight the importance of constructing robust and reliable void catalogs capable of accurately characterizing the underdense component of the cosmic web.
However, unlike overdense structures such as halos or galaxy clusters, voids do not possess a uniquely defined boundary or a universally accepted physical definition.
As a consequence, different void-finding methodologies often identify distinct structures depending on the adopted tracer distribution, density thresholds, topological criteria, or geometrical assumptions \citep{2008Colberg,comparison1,comparison2}.
This intrinsic ambiguity has led to the development of a broad variety of void identification algorithms, each optimized for different scientific goals and observational conditions \citep{2007.Platen.WVF,2008.neyrinck.zobov,2015.sutter.vide,2023.Correa.Voids}.

A widely used approach is based on the analysis of the matter density field \citep{2007.Platen.WVF, 2008.neyrinck.zobov, 2015.sutter.vide}.
By drawing an analogy between the density field and a topographic landscape through watershed transforms, these methods define voids as the basins that would be flooded by a progressively rising water level.
This family of void finders has the advantage of naturally recovering the complex morphology of cosmic voids without imposing a predefined shape, while also allowing the identification of hierarchical substructures within larger underdense regionsim

A contrasting approach consists in directly searching for regions whose integrated density contrast falls below a prescribed threshold \citep{1997El-Ad,2002Hoyle,2015.ruiz.cluesiii, 2019.ruiz.structure, 2016.banerjee.simulating}.
This family of methods define a template shape (generally spherical or quasi-spherical geometries) and convolve it on the matter or tracer density field until the enclosed region satisfies the underdensity criterion.
The simplicity of this geometrical description provides more direct control over the density properties of the identified voids, while also facilitating the interpretation of their effective sizes and volumes.
Due to the integrated density contrast being defined at large scales and being closely related with the theory of void abundances \citep{2013Jennings,2004.Seth.Hierarchy}, no filtering criteria is required for this kind of void finders \citep{2023.paz.popcorn}.
Nevertheless, the use of predefined geometrical templates may limit the capability of these methods to accurately recover the highly irregular morphology of realistic cosmic voids.
Recent approaches have addressed this limitation by allowing spherical regions to be combined into larger, non-spherical structures.
Earlier examples include \pvfl, which iteratively grows and combines underdense spheres to construct free-form void regions that jointly satisfy an underdensity criterion \citep{2023.paz.popcorn}, and the more recently developed VERSUS, which employs a merging criterion to identify nested void-like structures \citep{Findlay2026}.
In particular, \pvfl provides a geometrical extension of spherical void-finding algorithms that can recover highly non-spherical void regions while preserving computational simplicity and geometrical interpretability.

The increasing scale and complexity of cosmological simulations and observational datasets introduce significant computational challenges for astronomy \citep{2011.Berriman.Tsunami, 2017.Garofalo.Data}.
As modern analyses require the processing of increasingly larger tracer catalogs and higher-resolution density fields, computational performance, scalability, and efficient memory usage become critical aspects in the practical application of void identification methods.
At the same time, the growing complexity of scientific software has highlighted the importance of maintainable and extensible codebases, capable of supporting long-term development, reproducibility, and the integration of new analysis strategies \citep{2025.Buchner.Impact, 2019.Tollerud.Sustaining}. 
However, these aspects are often only briefly discussed in the original presentations of many void-finding algorithms.

In this sense, the initial implementation of \pvfl was more akin to a \textit{software prototype} \citep{1999.Hunt.Pragmatic}, aimed at demonstrating the viability of Popcorn voids as a geometrical extension of spherical voids.
While this approach successfully established the foundations of the method, the subsequent development resulted in a codebase containing obsolete implementations, experimental remnants, and little error handling.
The lack of a standardized usage procedure and comprehensive user documentation made it difficult to determine the intended way of configuring and running the software.
In this work, we present a complete reimplementation of \pvfl focused on modern software design, computational performance, parallel execution, and algorithmic extensibility.
Special attention is given to the detailed description of its algorithmic structure and implementation strategy, providing a clearer and more reproducible characterization of the methodology than previously available.
The resulting framework is designed to facilitate future developments, comparative studies, and large-scale cosmological analyses while maintaining a flexible and maintainable software architecture.

This work is organized as follows: in \autoref{sec:pop_algo} we formally introduce the \pvfl and describe the complete popcorn identification process.
In \autoref{sec:idc} we focus on the challenges arising from computing the integrated density contrast.
In \autoref{sec:ooopt} we introduce a set of heuristic optimizations that operate under relaxed constraints tailored to our specific cosmological scenarios.
Next, in \autoref{sec:arch} and \autoref{sec:res}, we describe the software architecture of \pvfl, with emphasis on its parallel model and replicability, and evaluate the contributions of this work through quantitative comparisons against the original software, respectively.
In \autoref{sec:concl} we wrap up with conclusions and future work.

\section{Popcorn voids in the Large-Scale Structure}
\label{sec:pop_algo}

Spherical voids operate under the assumption that voids can be approximated, in general terms, as underdense spherical regions.
Thanks to their straightforward and well defined properties, they have been successfully used to study the statistical properties of void populations \citep{2013.Paz.Cluesii, 2015.ruiz.cluesiii,2016Ceccarelli,2017Lares,2019Correa,2023.Correa.Voids} as well as the evolution of galaxies in these extremely low density environments \citep{2019.ruiz.structure,2020Alfaro,2022Alfaro,2022Rodriguez,2023Rodriguez}.
However, its main limitation lies in that very same assumption of spherical symmetry, which fails to capture the complex and irregular morphology observed in realistic cosmic voids.
As a result, spherical methods can fragment extended underdense regions into multiple smaller voids or fail to fully describe their geometry \citep{2023.paz.popcorn}.

\pvfl (\pvf hereafter) builds upon the spherical approach by relaxing the assumption of a fixed geometry.
Instead of restricting voids to single spheres, the algorithm defines them as connected unions of overlapping spherical regions by imposing to the joint volume the same underdensity criterion.
This construction allows voids to grow iteratively and adapt to the underlying matter distribution, effectively producing free-form structures while preserving a well-defined density-based selection criterion.
In this sense, popcorn voids can be understood as a natural generalization of spherical voids.

The identification of popcorn voids is inherently a multi-step process, combining physical assumptions, geometric constraints, and algorithmic design choices.
Rather than presenting the method as a monolithic procedure, we decompose it into its main components and describe the role played by each stage in the full pipeline.
This process is organized in two core algorithms.
The first one identifies spherical voids, which provide the initial seeds for the subsequent popcorn growth stage; and the second one attempts to expand these seeds into extended structures by iteratively adding spherical members while preserving the integrated density contrast criterion.
Both stages are controlled by the same set of parameters: the minimum and maximum search radii, $[R_{\min}, R_{\max}]$, and the integrated density contrast threshold for voids $\Delta_v$.

\subsection{The spherical void algorithm}

The spherical void finder works by discretizing the continuous volume into a regular cubic grid.
Beyond enabling the use of \nnnl algorithms for acceleration, each cell will also serve as initial candidates from which potential spherical voids will form.
The procedure, summarized in \autoref{alg:svf}, follows by applying a three-step process independently on each of those cells, producing at most one void per candidate region.

\begin{algorithm}
\caption{Spherical void finding process: divide the continuous volume in candidate cells and try to fit an underdense sphere on each one.}
\label{alg:svf}
\begin{algorithmic}[1]
\Require Search Bounds $[R_{\min},R_{\max}]$, Density Threshold $\Delta_v$
\Ensure Spherical voids
\State Build grid $G$

\For{each grid cell $g_c \in G$}
    \State Estimate candidate radius $r_i$

    \While{$r_i \ge R_{\min}$}
        \State Generate local centers $C$ inside $g_c$
        \State Reduce search range $[R_{\inf},R_{\sup}] \gets$ Bound($R_i$)
        \State Voids $\gets$ \{ FindVoid$(c_i,R_{\inf}, R_{\sup},\Delta_v)$ for $c_i \in C$ \}

        \If{Found valid void}
            \State Accept $\max\limits_{v \in \text{Voids}} v.\text{radius}$
            \State \textbf{break}
        \EndIf

        \State Decrease candidate radius $R_i$
    \EndWhile
\EndFor
\end{algorithmic}
\end{algorithm}

\textbf{Candidate radius estimation}:
Before performing a detailed search, find the largest void radius that the current cell could host.
This is done by computing quick lower-bound approximations of the density using discrete concentric spheres within the cell center.
The search begins at $R_{\max}$ and decreases until reaching a point below $R_{\min}$ by multiples of $R_{\max} / N_\mathrm{step}$, where $N_\mathrm{step}$ is the number of discretization steps.

\begin{equation}
    N_\mathrm{step} = \left\lceil \frac{R_{\max}}{R_{\min}} \right\rceil \text{.}
\end{equation}

Within a fixed discrete radius $r_i$, a lower-bound of the mass $M_\mathrm{inf}$ is obtained from a specific subset of tracers that are inside all possible sphere centers within the cell, and can be quickly determined without direct distance computations.
Then, the low estimated density is computed by:

\begin{equation}
    \rho_\mathrm{inf} = \frac{3 M_\mathrm{inf}}{4 \pi r_i^3} \text{.}
\end{equation}

If at no point was the density threshold satisfied, the cell is discarded; otherwise, we proceed to the next stage, recording $r_i$ as the initial size estimation for the potential spherical void.

\textbf{Local optimization}:
At this point, we can begin a detailed search of voids.
Using the discrete radius from the previous step (and its successor), we can derive a reduced range $[R_{\inf},R_{\sup}]$ in which we'll limit our refined search.

This process begins by generating a fixed number of random centers within the volume of the cell, then performing a bounded search procedure on each of them: for each center $\vec c$, scan for every tracer $\vec t$ located within a distance of $[R_{\inf},R_{\sup}]$, compute the exact spherical density centered at $\vec c$ with radius $||\vec t - \vec c||$.
Record the largest distance that satisfies the density threshold $\Delta_v$.
Overall, this results in a set of potential voids rooted in each random center.

\textbf{Progressive radius backtracking}:
If a subset of valid voids were found in the previous step, accept the one with the biggest radius and stop the procedure for this cell; otherwise, decrease the initial discrete radius $r_i$, and retry the local optimization.
If the upper radius bound goes bellow $R_{\min}$, stop the procedure without accepting any voids.

\subsection{Overlapping spherical voids}
\label{subseq:sph_overlap}

Since the spherical void finder is applied independently to every candidate cell, neighboring candidates frequently converge to nearly identical spherical voids associated with the same underdense region.
Consequently, the raw output contains a large number of highly overlapping spheres, requiring a dedicated post-processing stage to resolve these redundancies.

An important consideration is that non-overlapping sphere configurations are sensitive to the order of conflict resolution.
In our case, to preserve the largest sphere representing each void, these resolutions must be done in strictly decreasing order of radius.
At each iteration, the currently remaining sphere is selected as the primary candidate, its overlapping neighbors are identified, and the conflicts are resolved by selectively removing intersecting spheres.
From this process, we deduce the following invariant: any sphere that survived until its turn is necessarily the largest within its overlap set; otherwise, it would have already been removed in a previous iteration.
Consequently, size comparisons become unnecessary, and only spheres whose centers lie within a distance $2R_v$ (where $R_v$ is the radius of the current candidate) need to be considered during neighbor search.
Discarded spheres are excluded from subsequent overlap resolutions.

Overall, this procedure ensures that when multiple overlapping candidates represent the same vicinity, only the largest spherical void is preferentially retained.
However, this guarantee doesn't hold for voids in general.
In practice, physical voids are broken up into non-overlapping spheres that were unable to grow large enough because of their complex geometry that cannot be captured with a spherical shape.
These are the issues that \pvf was designed to address.

\subsection{The \textit{Popcorn} void expansion algorithm}

Once a spherical void catalogue is produced, the popcorn building procedure can begin.
This growth process, summarized in \autoref{alg:pvf}, is applied independently to every void in the set.
Rather than searching for an arbitrary three-dimensional boundary, the algorithm constructs a popcorn void as the union of spherical members.
At each iteration, it identifies the locally optimal sphere that can be attached to the existing structure while preserving the integrated density threshold.
This process works as follows:

\begin{algorithm}
\caption{Popcorn void identification: starting from a set of spherical voids, apply an iterative expansion process to each sphere, converting them into a set of popcorn voids.}
\label{alg:pvf}
\begin{algorithmic}[1]
\Require Spherical voids $S$
\Ensure Popcorn voids

\For{each root sphere $s \in S$}
    \State Initialize Popcorn void Pop $\gets \{ s \}$
    \Repeat
        \State Seed surface with Seeds $\gets$ FibonacciSeed(Pop)

        \While{Seeds is not empty}
            \State Cands $\gets$ \{inflate(Pop,seed) for seed $\in$ Seeds\}

            \State BestCandidate $\gets \max\limits_{c \in \text{Cands}} c.\text{volume}$

            \If{BestCandidate is too small}
                \State \textbf{break} \Comment{Stop Popcorn formation}
            \EndIf
 
            \State Pop.append(BestCandidate)
            \State Seeds.remove(BestCandidate.seed)
        \EndWhile
    \Until{No sphere was added}

    \State Accept Pop
\EndFor
\end{algorithmic}
\end{algorithm}

\textbf{Popcorn void initialization}:
Each popcorn void is initialized with a single spherical member, corresponding to the root sphere.
At this stage, the popcorn structure is equivalent to a spherical void and already satisfies the density threshold by construction.

\textbf{Surface sampling}:
To expand the void, the algorithm first identifies a set of candidate growth locations on the boundary of the current popcorn structure.
These locations, referred to as \textit{seeds}, represent potential centers from which new spherical members can be grown.

An adequate distribution of seeds is essential to ensure that all possible growth directions are explored while avoiding preferential orientations or poorly sampled regions of the void surface.
To this end, the surface of every spherical member is sampled using a Fibonacci lattice, which is a pattern widely observed in nature \citep[e.g. sunflower seeds, pineapple scales; see][]{1982.Ridley.sunflowers, 2009.Gonzales.meassurement}.
Fibonacci lattices provide a nearly uniform distribution of points over the sphere with optimal packing efficiency.
Furthermore, they admit arbitrary choices for the total number of points, allowing the sampling density to scale naturally with the
sphere radius while maintaining a nearly constant spatial resolution.

These properties make Fibonacci lattices particularly well suited for the iterative growth stage, where an isotropic and homogeneous coverage of the popcorn boundary is required.

\textbf{Local growth of satellites}:
For each surface seed, attempt to append a new spherical member by growing an underdense sphere around the seed position.
This process mirrors the spherical void finder (without recentering) and is carried out in two steps:

\begin{enumerate}
    \item A preliminary filtering stage estimates whether a sphere centered at the seed could satisfy the integrated density contrast threshold for the joint volume.
    \item If successful, a bounded refinement step determines the maximum admissible radius for such sphere.
\end{enumerate}

During this process, the density evaluation accounts for the tracers already contained in the popcorn structure, ensuring that the union of spheres remains underdense.
Each successful inflation produces a candidate spherical member with a well-defined radius, volume, and density contrast.

Among all inflated candidates, we select the sphere that contributes the largest additional volume to the popcorn structure.
If the radius of this sphere falls below the minimum allowed scale, the growth process is halted.
Otherwise, the selected sphere is added to the popcorn void, and the corresponding seed is removed from the candidate set.
The remaining seeds are retained, allowing the algorithm to explore alternative growth directions in subsequent iterations.

This greedy strategy prioritizes the most significant geometric expansions while maintaining computational efficiency.

\textbf{Iterative grow and termination}:
The surface sampling and growth process are repeated iteratively until no additional spherical member can be grown.
When no seed yields a valid sphere satisfying the density threshold, the popcorn void is considered complete.
Notably, if no additional spheres could be added, the initial root sphere itself constitutes a valid popcorn void.

\subsection{Overlapping popcorn voids}
\label{subseq:pop_overlap}

A direct consequence of the fragmentation problem in spherical void catalogs, is that the popcorn voids that grew from fragmented voids produce almost identical overlapped structures.
This behavior is expected, otherwise it would mean the expansion failed to correctly capture the geometry of the void.
Consequently, it is easy to detect and remove the redundant voids, as they will have relatively big overlapping ratios.

The first step of the cleanup procedure consists of identifying all overlapping popcorn pairs.
Two popcorn voids are considered overlapping if at least one spherical member of the first structure overlaps with one spherical member of the second.
For every overlapping pair, we compute the volume of the union formed by all spherical members of both popcorn voids.
The intersection volume is then obtained from the inclusion/exclusion relation
$$
V(A \cap B) = V(A) + V(B) - V(A \cup B),
$$
where $A$ and $B$ denote the two popcorn voids.
Afterwards, each pair is resolved individually.
If none of the voids from the pair were previously removed, and the overlapping volume is greater than the minimal sphere, i.e., $V(A \cap B) > (4/3) \pi R_{min}^3$, the voids are considered ``equivalent'' and a filtering criterion is applied.

The straightforward choice for a filtering criterion is to keep the void with largest volume, mirroring the greedy strategy adopted by the spherical void finder.
Instead, \pvf retains the void whose root sphere has the largest radius.
This choice is motivated by the observation that larger root spheres typically occupy the central region of an underdense structure, whereas smaller root spheres are more likely to represent peripheral fragments of the same void.
Consequently, popcorn voids grown from larger roots tend to capture the global structure of the underdense region, while those grown from smaller roots predominantly reproduce it from the outside in.
This behavior also motivates the early growth termination criterion discussed in \autoref{subseq:convergence}.

So far we have covered the construction process of a popcorn void from a tracer set of the matter distribution, where the core criterion governing void identification is the integrated density contrast within a candidate region.
In the following section, we shift our focus to this criterion, along with the challenges imposed by very large catalogs and complex geometries.

\section{Integrated density contrast computation}
\label{sec:idc}

In the study of the large-scale structure of the Universe, the matter distribution is commonly characterized through the density field, where we define the density contrast $\delta(\mathbf{x})$ \footnote{With an implicit time dependency considered constant for our purposes.} as local deviation of the matter density $\rho(\mathbf{x})$ with respect to the mean cosmic density $\bar \rho$:

\begin{equation}
    \delta(\mathbf{x}) =
        \frac{\rho(\mathbf{x})}{\bar \rho} - 1
    \, \text{.}
\end{equation}

Within this framework, overdense regions correspond to locations where
$\delta(\mathbf{x}) > 0$, while underdense regions satisfy
$\delta(\mathbf{x}) < 0$.
Cosmic voids emerge as extended connected regions whose matter content is significantly below the cosmic mean density, representing the most underdense environments of the cosmic web.

Since large scale structure evolves from initially small density perturbations due to gravitational instability, the integrated density in a given region is what defines if a region is going to expand or collapse with respect to the expanding universe background metric \citep{1980.Peebles.Lss}.
Voids are underdense regions that are expected to expand becoming more underdense as time goes, therefore, the characterization of voids is usually formulated in terms of an integrated (averaged) density contrast over a finite volume $V$.
Formally, a region can be considered void-like when its enclosed density contrast $\Delta$ falls below a prescribed threshold $\Delta_v$:

\begin{equation} \label{eq:big_delta}
    \Delta = \frac{1}{V} \int_V \delta(\mathbf{x}) \, d^3x
    = \frac{M / V}{\bar \rho} - 1
    \le \Delta_v
    \, \text{,}
\end{equation}
where $M$ is the total mass within $V$.

Consequently, computing the integrated density contrast of a candidate reduces to computing mass and volume, both of which involve special performance and geometrical considerations.
We begin with the mass accumulation, which heavily relies on spatial structures to avoid traversing the entire simulation for particle queries.

\subsection{\nnnl}

Efficient evaluation of local tracer densities requires restricting distance computations to a limited spatial neighborhood.
In \pvf these evaluations are accelerated through a regular spatial grid that enables efficient \nnnl (\nnn) queries.
By partitioning the search space into cubic cells, distance computations can be restricted to a finite neighborhood around each point of interest, avoiding unnecessary evaluations over distant tracers that cannot contribute to the local density estimate.
This strategy is equivalent to the linked-cell method commonly used in particle and fluid dynamics simulations, which guarantees completeness of neighbor searches within a finite radius \citep{1988.Hockney.Particles, 2017.Allen.Liquids}.

Beyond accelerating \nnn queries, the same grid is used to define the candidate centers explored by the spherical void finder.
This dual role introduces a competing design requirement: on one hand, the cell size should be large enough to maximize the efficiency of neighbor searches; on the other, it must provide sufficient spatial resolution to sample the large-scale structure without missing potential void candidates.
To balance these constraints, the grid resolution is determined by combining a density-based criterion with a geometric constraint: begin by choosing an initial guess for the number of cells, and then refine it if the cells are too large.
Consequently, this constraint will override the initial density-based estimate most of the times, a limitation we revisit in \autoref{subseq:decoupling}.

\begin{figure}
    \includegraphics[width=0.98\columnwidth]{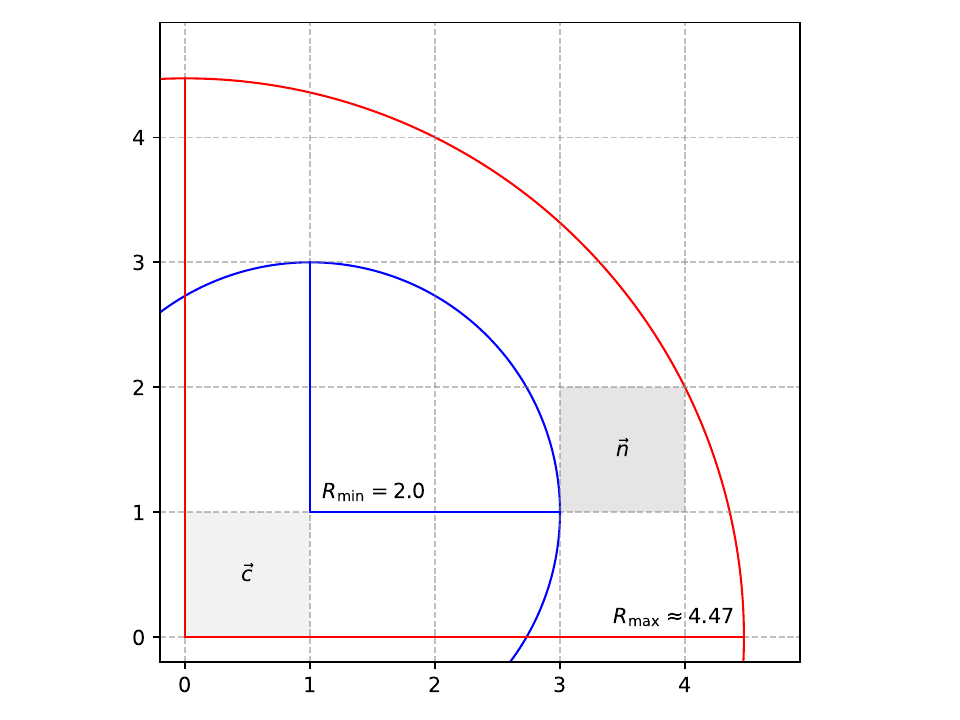}
    \centering
    \caption{2D representation of the neighbor overlapping algorithm.
    Given a neighboring cell $\vec{n}$; $R_{min}$ and $R_{max}$ represent the \textit{closest-smallest} and \textit{furthest-biggest} spheres that make contact with $\vec{n}$, originated in the central cell.}
    \label{fig:overlap}
\end{figure}

In order to minimize distance computations, we use an algorithm that can quickly determine if a neighboring cell is outside, inside, or partially contained within a candidate sphere.
This allows us to directly discard or accumulate all the objects on cells in the optimal cases, falling back to computing particle-by-particle distances for cells in contact with the shell of a candidate sphere.
The idea behind this procedure is that inside a central cell $\vec c$, there's a minimum radius sphere that would never cover a neighboring cell regardless of it's position inside $\vec c$.
An analogous case can be derived with a maximum radius sphere that can always cover a neighboring cell.
\autoref{fig:overlap} illustrates the geometrical meaning of these criteria using a two-dimensional example.

During popcorn void expansion, candidate spheres are appended to an existing structure that contains an invariant set of particles with a cumulative mass, regardless of the size and position of such candidates.
To avoid redundant recomputation of these invariant quantities, a particle vector is maintained for the incremental popcorn structure.
Additionally, a mask is introduced to prevent the same object from being considered twice during accumulations, which happens for particles in the shared volume between the growing popcorn structure and the candidate.

After determining the mass contained in our region of interest, we just need its volume to compute the integrated density contrast.
For spherical regions, this is easily determined in terms of its radius.
Popcorn voids, however, don't have a straightforward method to obtain their volume, which is the next problem we will tackle.

\subsection{Volume of a popcorn void}

Popcorn voids are generally composed of many partially overlapping spheres.
Consequently, computing the volume of such structures requires evaluating the volume of the union of multiple spherical regions.
Analytical expressions exist for simple configurations involving two or three spheres, obtained by summing individual volumes and subtracting the overlapping regions (spherical caps) \citep{1964.Bondi.2sph, 1982.Pavani.3sph}.
However, these expressions quickly become impractical for larger systems.
In the general case of many spheres we employ the approach presented in the \textsc{ARVO} package \citep{2005.Busa.Arvo, 2012.Busa.ArvoCL}.

This technique numerically computes the volume of the union of $N>2$ spheres,
$$
M = \bigcup_{j=1}^{N} S_j ,
$$
by transforming the volumetric integral into a set of surface integrals over the spherical boundaries that remain exposed after accounting for overlaps.
Formally,

\begin{align*}
V(M) &= \iiint_{V(M)} dx\,dy\,dz \\
     &= \iint_{B(M)} z\,dx\,dy \\
     &= \sum_{i=1}^{N} \iint_{B_i(M)} z\,dx\,dy ,
\end{align*}
where $B(M)$ denotes the boundary of the union of spheres and $B_i(M)$ represents the portion of the surface of sphere $S_i$ that is not occluded by any other sphere.

\begin{figure}
    \includegraphics[width=0.96\columnwidth]{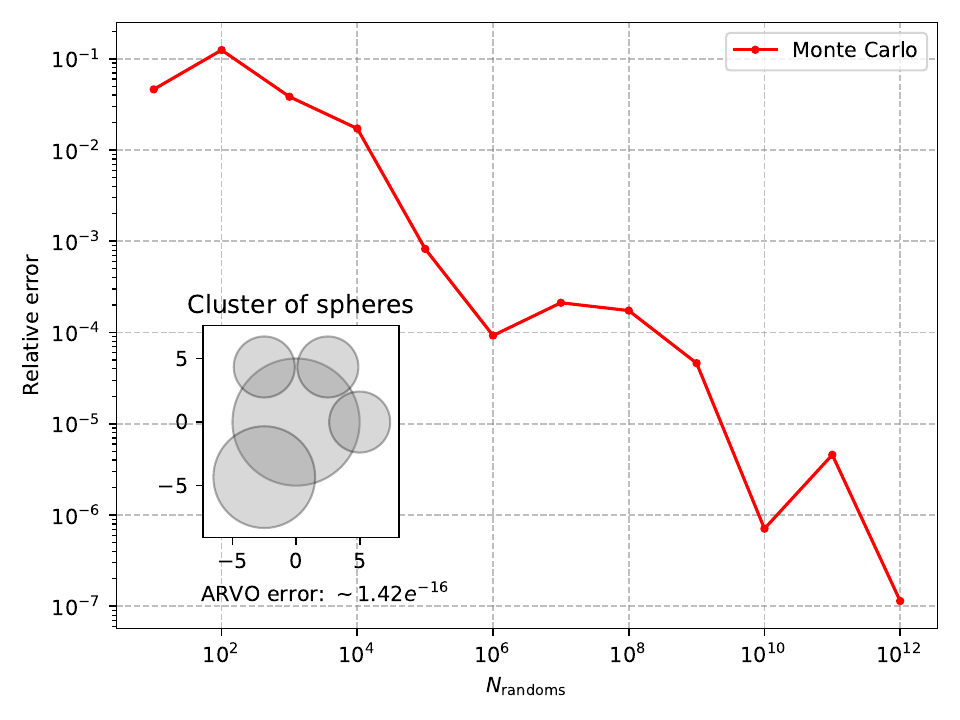}
    \centering
    \caption{Monte Carlo error in terms of the number of randoms computing the volume of the cluster of five overlapping spheres in the bottom left panel.}
    \label{fig:mc_vs_arvo}
\end{figure}

To evaluate the numerical behavior of the adopted volume computation, we designed a controlled configuration composed of five overlapping spheres, in a way such that no more than two spheres overlap.
In this scenario, the exact union volume can be computed as the sum of the individual sphere volumes minus the corresponding pairwise spherical-cap contributions.
This value served as a reference to compare ARVO against a Monte Carlo estimator using $N=10^k$, with $k = {1,2,\ldots,12}$ random points.
In \autoref{fig:mc_vs_arvo}, we can see that the Monte Carlo estimates improved gradually, reaching an error of order $10^{-7}$ with $N=10^{12}$ randoms.
On the other hand, ARVO achieved a relative error of order $10^{-16}$.
The achieved accuracy, paired with the deterministic nature of the algorithm, makes ARVO an effective method for computing the volumes of popcorn voids.

\section{Heuristic optimizations}
\label{sec:ooopt}

So far, we have covered a baseline solution utilizing standard generic optimizations.
These conventional techniques maintain strict mathematical correctness and uniform reliability across all general input.
However, cosmological structures present unique statistical features that allow for further efficiency gains.
In this section, we introduce two pragmatic optimizations that exploit the statistical properties of our target cosmological datasets.
These methods introduce a relaxed coherence model under specific conditions in exchange for significant performance gains.

\subsection{Early growth termination}
\label{subseq:convergence}

As popcorn voids grow from the underdense spheres, fragmented regions that were initially represented by multiple spherical voids often end up with the same number of equivalent popcorn voids.
While these redundancies are resolved during the final cleanup stage, a substantial amount of computation may have already been spent growing these structures that will ultimately be discarded.
To circumvent this problem, following the initial publication and the first release, optimizations to the legacy code were tested within development versions by implementing an early termination of growth in the Popcorn algorithm.

Let $V_n$ denote the popcorn volume after the $n$-th spherical member has been added.
We define the relative volume increment:

\begin{equation}
    \delta_n = \frac{V_n - V_{n-1}}{V_n} \, \text{,}
\end{equation}
which measures the fractional contribution of the newly added sphere to the current popcorn volume.
During the growth process, the termination criterion enforces the sequence $\{\delta_n\}$ to decrease monotonically.
Whenever a newly added sphere produces a larger relative increment than the previously accepted member, the expansion is terminated early.

Complementary to the overlapping popcorn filtering procedure discussed in \autoref{subseq:pop_overlap}, the rationale behind this criterion stems from the way the spherical void finder samples underdense regions.
By construction, the spherical catalog is composed of maximal non-overlapping spheres, so a typical void is represented by a small number of large spheres occupying its central region, together with several progressively smaller spheres filling the remaining space toward the void boundaries.
Consequently, when a popcorn void is grown from one of these central spheres, each newly added member is expected to contribute a progressively smaller fraction of the total volume as the structure adapts to the surrounding geometry.
Conversely, in the smaller satellite spheres, the first expansion steps tend to produce increasingly larger members as the algorithm approaches the centre of the underdense region.

\begin{figure}
    \includegraphics[width=0.98\columnwidth]{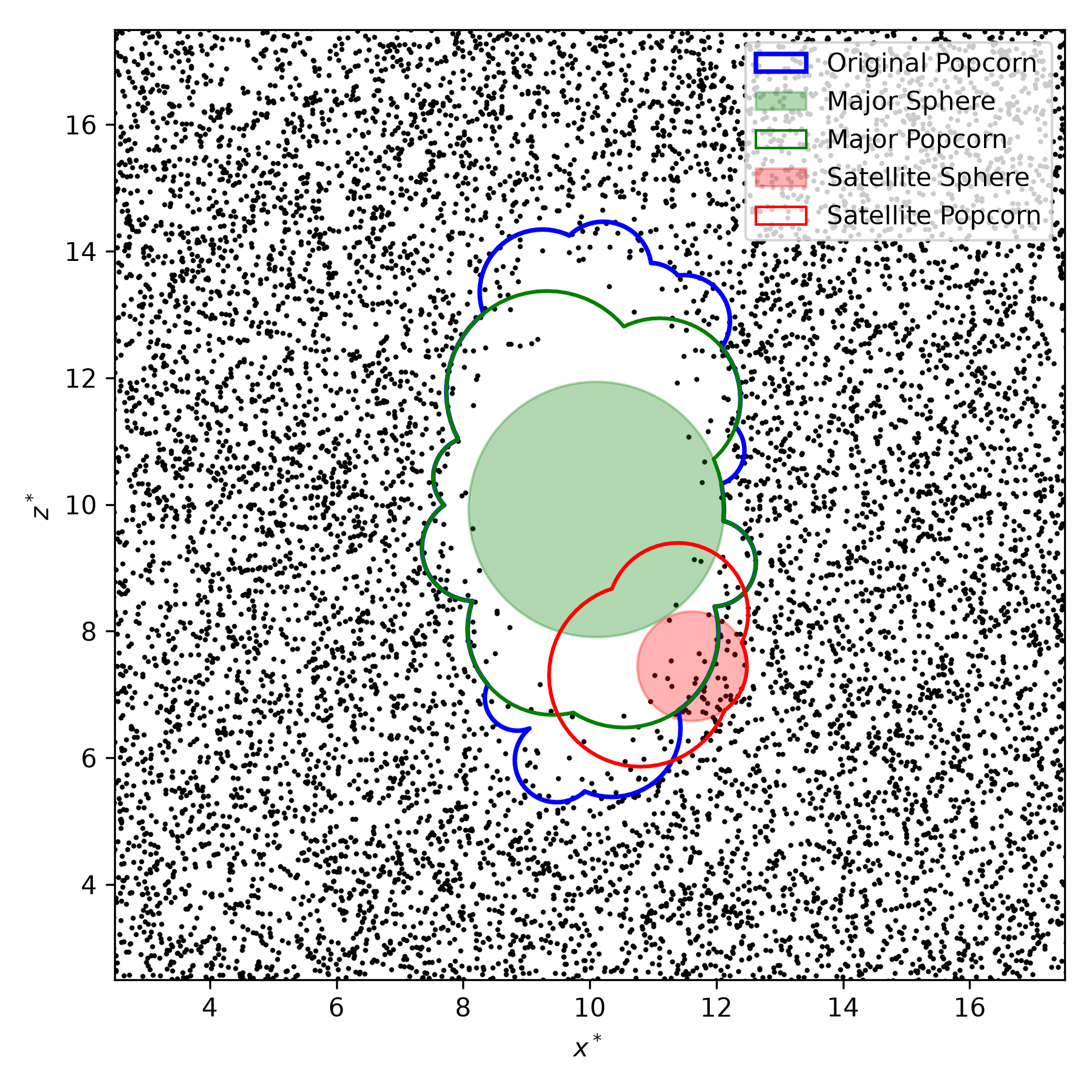}
    \centering
    \caption{Illustration of the popcorn early termination criterion on a synthetic example, with the original popcorn void plotted as a blue outline.
    In green we represent the biggest (major) spherical void found in the region, along with the resulting popcorn void obtained with this heuristic.
    Similarly, in red, we represent one of the satellite spheres with its respective popcorn void.
    }
    \label{fig:pop_heuristics}
\end{figure}

We can observe this criterion in action in \autoref{fig:pop_heuristics}, an underdensity with an elongated geometry that is poorly represented by spherical voids.
In this example, the spherical void finder identified seven spheres, but for visualization purposes we focused on two of them: a major sphere and one of its satellites.
Under normal conditions, the algorithm expands all spheres into very similar extended shapes, with a maximum volume difference of $4.3\%$.
With the heuristic in place, the difference becomes more significant: the major sphere became a popcorn with ten members, while the satellite only reached three; overall, the volume decreased by $75.9\%$.
However, this change introduced an unintended regression: the surviving popcorn void is significantly smaller than the original ($16.0\%$ reduction in volume), showing that this monotonic criterion negatively affects the growth of the voids that effectively survive the cleanup stage.

Despite the degradation in size, this heuristic substantially reduced the amount of work required to construct the voids.
To circumvent this, we extended the criterion by allowing a small number of ``infringements'' to the strictly decreasing sequence introduced in the legacy criterion.
The key lied on defining this number of infringements in terms of the size of the root sphere, making this criterion aggressively strip smaller voids, while loosely allowing the bigger voids to grow.
Overall, this adaptive criterion was able to recover the growth behavior observed in the original popcorn algorithm for the synthetic example shown in \autoref{fig:pop_heuristics}, while retaining the computational benefits of early termination.
Nevertheless, it remains inherently susceptible to prematurely terminating the growth of complex structures in certain configurations.
The resulting trade-off between computational performance and statistical properties are evaluated in more detail in \autoref{sec:res}.

\subsection{\nnn-candidate decoupling}
\label{subseq:decoupling}

As we previously established, a core component of \pvf is the spatial grid, which serves as an acceleration structure, as well as the initial candidate set for spherical voids.
This dual role introduces potentially competing design requirements.
The candidate cell must be fine enough to sample the search space without missing potential voids, but coarse enough to avoid generating redundant candidates.
In contrast, the \nnn cell should be sized to maximize query efficiency by balancing the number of tracers per cell against the effectiveness of the local search optimizations.
To address this potential issue, we explored an alternative design in which the \nnn grid would be independent from the candidate centers.
Here, we would keep the original candidate resolution and search for optimal \nnn grid parameters based on input conditions.

The \nnn resolution can be determined through the target mean number of particles per cell, denoted by $N_0$.
This parameter is related to the number of grid cells per dimension through:

\begin{equation}
    N_\mathrm{grid} \approx
    \mathrm{ceil}\left[ \left(\frac{N_\mathrm{part}}{N_0}\right)^{1/3} \right]
    \text{,}
\end{equation}
where $N_\mathrm{grid}$ is the number of cells per dimension on the grid, and $N_\mathrm{part}$ is the total number of particles.

\begin{figure}
    \includegraphics[width=0.98\columnwidth]{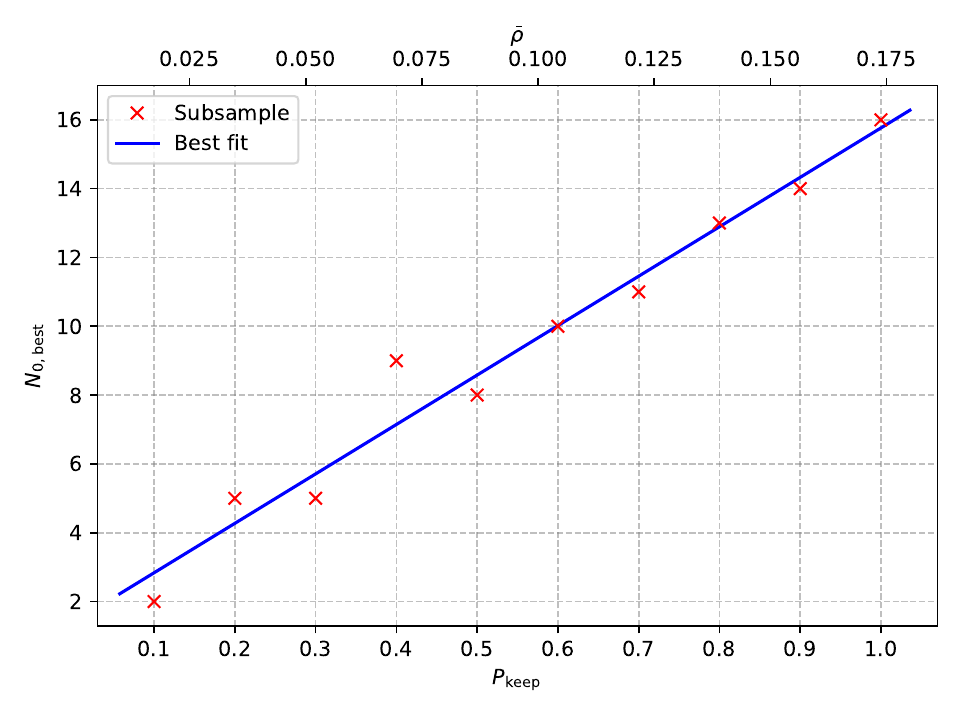}
    \centering
    \caption{Relationship between the mean particle density $\bar{\rho}$ and the optimal mean number of particles per cell $N_{0,\mathrm{best}}$ in a \nnn grid structure on random sub-samples of a simulation, discarding particles with $P_\mathrm{keep} = \{1.0, 0.9,\ldots,0.1\}$.
    }
    \label{fig:n0_fit}
\end{figure}

Selecting an appropriate value for $N_0$ is critical, as excessively small cells increase indexing overhead, while excessively large cells degrade neighbor-search locality.
To determine these suitable values, we performed an empirical study using random sub-samples of a simulation with a cubic volume of $1000h^{-1}\mathrm{Mpc}$ \footnote{In astronomy, cosmological distances are usually expressed in Megaparsecs, where $1~{\rm Mpc} = 10^6~{\rm pc} = 3.08568 \times 10^{19}~{\rm km}$ \citep{allen_2000}. It is also a common practice the use of $h^{-1}{\rm Mpc}$, where $h$ parametrizes determines the value of the present- day Hubble parameter as $H_0=100h~{\rm km~s}^{-1}{\rm Mpc}^{-1}$, a convention practice inherited from the first galaxy redshift surveys \citep{cfa_survey_1983,las_campanas_1997}.} on a side and containing approximately $174$ million particles.
Each sample was generated by dropping particles with a random probability $P_\mathrm{keep} = \{1.0, 0.9,\ldots,0.1\}$.
For each case, we evaluated multiple discrete values of $N_0$ and selected the configuration yielding the lowest execution time.
The resulting empirical relation as a function of the mean density $\bar \rho$ is shown in \autoref{fig:n0_fit} and is approximated by:

\begin{equation}
\begin{aligned}
    N_0 &= a \bar \rho + b &
    a &\approx 82.656 &
    b &\approx 1.4
    \text{.}
\end{aligned}
\end{equation}

This empirical relation, however has a limited range of applicability.
At sufficiently high tracer densities, increasing the cell resolution eventually becomes counterproductive, as the cost of opening and processing individual cells is dominated by the distance evaluations within them.
Conversely, at very low densities, the optimal values fall below one tracer per cell on average, making $N_0$ an unsuitable parameterization of the search resolution.
Thus, the fitted relation should not be interpreted as a universal prescription, but as an empirical characterization of the regime explored in our simulations.

Despite these limitations, the experiment demonstrates the usefulness of treating candidate sampling and neighbor searches as separate design problems.
In particular, it shows that the optimal neighbor-search resolution is strongly dependent on tracer density and need not coincide with the resolution required to sample candidate void centers.
The current implementation therefore provides a first step towards independently tunable spatial discretizations, while the remaining limitations motivate a more general treatment of these scales in future developments.

\section{Software architecture}
\label{sec:arch}

The \pvfl pipeline is composed of three executable units:

\begin{description}
    \item[Spherical Void Finder (\texttt{pvf.svf})] identify spherical voids from a discrete tracer dataset.
    \item[Popcorn Void Finder (\texttt{pvf.popcorn})] grow spherical voids into popcorn voids with a discrete tracer dataset.
    \item[Cleanup (\texttt{pvf.cleanpop})] identify overlapping popcorn void pairs and remove the redundant voids.
\end{description}

All executables run with a single parameter file, specifying input tracer dataset, output void catalogs, and the finder parameters for the spherical and popcorn algorithms.
We support a wide range of input dataset formats coming from some of the most significant cosmological simulations \citep{2005.Volker.Gadget2, 2021.Volker.Gadget4, 2024.schaller.swift, HDF5}.

\begin{description}
    \item[\pvf formats] simple space separated ASCII, and a packed binary format inherited from the legacy code.
    \item[Particle catalogues] GADGET2 and GADGET4 binary formats, Gadget4 HDF5 format and Swift HDF5 format.
    \item[Group catalogues] GADGET4 subhalos and FoF groups, Swift FoF groups; all in HDF5 format. 
\end{description}

HDF5 support is disabled by default, as it is not essential to the finding process. Placing these features behind conditional compilation allows the code to be built without the HDF5 library. If a build without HDF5 support receives input in one of these formats, the program exits with an error message explaining how to enable the feature.
This feature, absent from the original codebase, was introduced to improve the user experience. As a result, the new codebase only requires a compiler supporting C++20 or later with OpenMP support, and the MPI library \citep{openmp, MPI}. Since parallel execution is essential for identification in large datasets, we next describe the role these libraries play in \pvf.

\subsection{Parallel execution model}

During the void identification process, the input dataset is treated as immutable, i.e., no modifications to the particle distribution occur while void candidates are being evaluated.
This enables the adoption of a simple parallel strategy across many levels.

The implementation adopts a hybrid parallelization model, combining distributed-memory parallelism via MPI with shared-memory multi-threading throug OpenMP (OMP hereafter).
At the highest level, the input space is partitioned into uniform blocks that are distributed among MPI ranks.
In the spherical void finder (\texttt{pvf.svf}), this corresponds to subsets of candidate cells, while in the popcorn stage (\texttt{pvf.popcorn}), the workload is distributed over the initial spherical voids.

Within each MPI process, the assigned workload is further subdivided among OMP threads, which independently evaluate candidate voids inside their respective subsets.
This hierarchical approach allows the algorithm to scale across compute nodes while maintaining efficient thread-level parallelism within each node.
The cleaning stage in SVF introduces a global synchronization constraint: it can only be performed once the identification process has been completed, and the full set of candidate voids is locally available.

From a runtime standpoint, the implementation allows flexible configurations of MPI ranks and OMP threads, enabling hybrid parallel execution across distributed and shared memory architectures.
Nevertheless, the current design exhibits an important limitation inherited from the legacy \pvf: each MPI rank maintains a full copy of the tracer catalog.
While this approach simplifies data access patterns and avoids communication overhead during the detection phase, it increases memory consumption due to redundant storage across ranks.
As a consequence, the implementation tends to favor configurations with a reduced number of MPI ranks and a higher degree of OMP threading per node.
We deliberately preserved this design choice to ensure methodological continuity with previous versions of \pvf and to facilitate future experiments we discuss in \autoref{sec:concl}.

\subsection{Reproducibility}

Besides reproducing scientific results, deterministic execution is particularly valuable during software development, as it simplifies debugging by allowing modified versions of the code to be compared against a known reference output.

In \pvf, spherical void identification relies on stochastic sampling which is executed through multiple layers of parallelism, making the sequence of random numbers dependent on runtime conditions.
The original implementation did not address this, accepting the fact that many runs on identical inputs would produce different outputs; often being noticeable due to the high sensitivity of spherical void sizes with their chosen center. Although this stochasticity does not affect the scientific validity of the results, it hinders a clear and unequivocal validation of the results during the development of the code.
So naturally, we introduced a deterministic execution mode that controls the sources of randomness and order of execution, ensuring reproducible results under fixed runtime conditions.

First, we replaced the base \texttt{C} pseudo-random generator with the \texttt{Xoshiro128*} float generator \citep{2018.Blackman.Xoshiro}, which is well known for good performance and high quality generation for general purposes.
More importantly, it can be separated into multiple independent streams.
At the MPI level, each process derives its own independent seed from a global seed through a deterministic transformation, so that all processes generate distinct, non-overlapping random sequences while preserving reproducibility across runs.
Within each MPI process, the OMP threads create and seed their own independent random streams trough a sequence-splitting strategy.

Beyond number generation, OMP scheduling policies introduce an additional source of randomness: with dynamic scheduling, iterations may be assigned to different threads across executions, changing the sequence in which thread-local random number generators are consumed.
Consequently, identical seeds no longer guarantee reproducibility.
A static scheduling policy eliminates this effect by assigning the same iterations to the same threads.
However, since the computational cost of the identification stages can differ across candidates, static scheduling could be prone to load imbalance.
To account for both scenarios, we opted for OMP's runtime scheduling instead.
This allows the user to make this choice according to their needs: static for reproducibility, or dynamic for performance.

The last source of variability is the final ordering of the voids. In the original implementation, voids were simply sorted by radius before writing the output. However, this leaves a subtle edge case: two voids of equal radius may appear in a different order across runs, which, although rare, occurred during testing. We therefore introduced the original candidate cell ID as a secondary sorting key, which breaks ties, guarantees a strict ordering, and fully resolves the reproducibility issue.

This approach has one limitation: reproducibility is guaranteed only for a fixed number of processes and threads and a fixed scheduling policy. This limitation is common in HPC codes (e.g., Sparkling, GADGET; \citealp{2015.ruiz.cluesiii, 2021.Volker.Gadget4}) and represents a typical trade-off for achieving deterministic execution without sacrificing performance or substantially changing the implementation.

\section{Results and comparison}
\label{sec:res}

To assess the impact of this work, we compared three aspects of both implementations: \textit{static analysis}, using software metrics to characterize the structural complexity and maintainability of the code; \textit{statistical efficiency}, evaluating physical results of a real cosmological scenario; and \textit{performance}, measuring the runtime behavior of the codes.
This benchmark evaluated the latest public release of the legacy Popcorn code \footnote{\url{https://gitlab.com/dante.paz/popcorn_void_finder}} (rev \texttt{63cbabe8}) against two tagged revisions of our implementation: \texttt{cpc-benchmarks} for statistical and performance measurements; and \texttt{cpc-final} for static analysis \footnote{These revisions differ only in their configuration file format, with negligible impact on performance or statistical results.}.

\subsection{Static analysis}

Static analysis provides a quantitative view of software structure and implementation complexity.
While this does not directly correspond with performance or functional soundness, it offers useful indicators of maintainability, readability, and overall design quality.
Given the data-centered nature of the code and the absence of class hierarchies and polymorphism, the analysis primarily focuses on metrics that characterize code size and control-flow.

\begin{figure}
    \includegraphics[width=0.99\columnwidth]{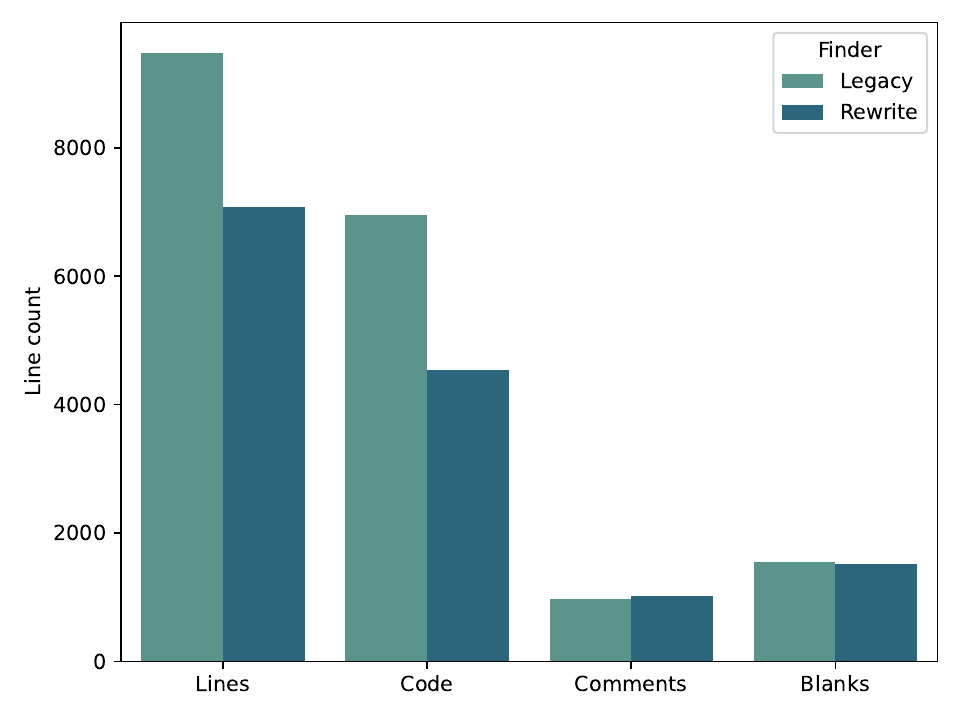}
    \centering
    \caption{Categorical distribution of Lines of Code (LOC) on the different finder implementations.}
    \label{fig:loc_overview}
\end{figure}

The first tool used in this analysis was \textit{Tokei} \citep{tokei}, which provides a categorical overview of the codebase in terms of the lines of code (LOC).
Tokei was executed over all project sources, including header files, documentation, implementations, and test suites; resulting in the metrics summarized in \autoref{fig:loc_overview}.
The most notable change can be seen in the reduction of the total line count, which decreased by approximately 25\%.
Looking more in depth, we can see that this was almost exclusively in code fragments, while comments and blank lines remained in similar amounts.
This was unexpected, as the new release provides detailed documentation on almost all public interfaces, unlike the legacy version.

While both implementations address the same problem using equivalent algorithms, their internal structures differ significantly.
A substantial amount of redesign work was carried out in the new release, which is not fully captured by the high-level overview provided by simple line-count metrics.
To gain further insight into the effects of this reengineering process, we made use of \textit{Lizard} \citep{lizard}, a tool focused on analyzing control-flow complexity through the \textit{Cyclomatic Complexity Number} (CCN) metric.
The main results are reported in \autoref{tab:ccn}.

\begin{table}[h]
\centering
\caption{Summarized metrics of Lines of Code (LOC) and the Cyclomatic Complexity Number (CCN) obtained from Lizard.}

\begin{tabular}{lrrrr}
\toprule
Finder  & Warnings & Functions & $\overline{\mathrm{LOC}}$ & $\overline{\mathrm{CCN}}$ \\
\midrule
Legacy  & 11 & 198 & 30.1 & 5.3 \\
Rewrite &  0 & 206 & 14.9 & 2.6 \\
\bottomrule
\end{tabular}
\label{tab:ccn}
\end{table}

While both implementations maintain a comparable number of functions, the new release reduces the average LOC and CCN by approximately half, suggesting a shift toward reduced complexity and greater conciseness.
Additionally, Lizard emitted 11 warnings in the legacy implementation, all of them on functions with $\mathrm{CCN} \ge 15$, which is generally considered as complex structure with moderate risk.
While some of these warnings refer to dead code, others are from fundamental functions in the identification process.
The diagnostics, along with a comparison against their new counterpart are presented in \autoref{tab:lizard_warnings}.

The most prominent case is seen in \texttt{inflate\_fast} which saw a decrease of roughly 90\% in both CCN and NLOC.
Here, the functionality was split across three new sub-processes:
\begin{itemize}
    \item Inflate a seed discretely using lower-bound $\Delta$ approximations
          ($\mathrm{NLOC}=48/\mathrm{CCN}=9$).
    \item Compute the the exact $\Delta$ of a seed with a fixed radius
          ($\mathrm{NLOC}=50/\mathrm{CCN}=8$).
    \item Find the exact void-boundary particle within a bounded range $[R_\mathrm{inf},R_\mathrm{sup}]$
          ($\mathrm{NLOC}=80/\mathrm{CCN}=13$).
\end{itemize}
Overall, this application of algorithmic decomposition and separation of concerns reduced the algorithmic complexity, and therefore risk, on the new release.

\begin{table}[]
\centering
\caption{Functions from the original implementation flagged for their high CCN ($\ge 15$) compared against their alternative in the new codebase.}
\begin{tabular}{lrrrr}
\toprule
& \multicolumn{2}{c}{NLOC} & \multicolumn{2}{c}{CCN} \\
\cmidrule(lr){2-3} \cmidrule(lr){4-5}
Function & Old & New & Old & New \\
\midrule
\texttt{inflate\_fast}      & 333 & 37 & 65 &  6 \\
\texttt{compute\_intersecs} & 163 & 53 & 24 &  6 \\
\texttt{svf\_work}          & 124 & 80 & 23 & 12 \\
\texttt{popcorn\_work}      & 122 & 84 & 18 & 13 \\
\bottomrule
\end{tabular}
\label{tab:lizard_warnings}
\end{table}

Lizard also provides an extension for detecting duplicated code, enabled through the \texttt{-Eduplicate} flag.
Using this feature, the analysis reports that $29.91\%$ of the legacy codebase consists of duplicated code segments.
Most of these duplicates are concentrated in three main areas: the finder algorithms, the input/output (IO) subsystem, and parsing logic.
A portion of the duplicated code within the finder algorithms corresponds to dead code.
For this reason, we focused on the latter areas.

Within the IO components, many sources of the duplication were found in shared logic for reading simulation snapshots split across multiple files, repeated implementations for handling the different variants of Gadget binary formats, and common boilerplate associated with HDF5 input routines.
In the new release, these patterns were addressed through more generic abstractions.
Multi-file snapshot reading is now implemented through a reusable scheduling routine that accepts a callable responsible for reading individual files, while managing resource allocation and execution flow.
Similarly, we implemented a unified Gadget reader that supports all binary formats through a parameterized function that allows format-specific behavior to be handled without duplicating the surrounding logic; which was possible because these only differ in the presence of headers in the datablocks.
The introduction of this abstraction, together with the adoption of the HDF5 Lite API, significantly reduced boilerplate code in the IO layer.

Additional duplication in the legacy code originated from repeated configuration parsing logic within the main executable sources.
This is a liability because any change in parameters must be applied repeatedly and consistently throughout all the files.
In the new release, configuration parsing is centralized into a dedicated module to ensure consistent behavior and facilitate future extensibility, such as introducing a validation step with detailed error descriptions.
As a result of these refactoring efforts, the duplicate code rate was reduced from $29.91\%$ to $2.68\%$ in the new release.

\subsection{Statistical efficiency}

We now turn to the runtime behavior of the code, running the complete pipeline on a realistic cosmological dataset to compare both the statistical results and the computational performance of the two implementations.
The simulation was carried out with GADGET2 \citep{2005.Volker.Gadget2}, which consisted of a flat $\Lambda$CDM universe in a $500 \ h^{-1}\mathrm{Mpc}$ box with $512^3$ tracers of dark matter, and the following cosmological parameters:

\begin{equation}
\begin{aligned}
    \Omega_{m,0} &= 0.258 & \Omega_{\Lambda,0} &= 0.742 & H_0 = 0.796  \text{.}
\end{aligned}
\end{equation}

Our proposed target for this test is the \vsfl (\vsf), which counts the number of voids of a specific radius per unit-volume in the universe.
As \pvf's main motivation is geometric refinement over the spherical void finders, this test provides the main highlight of popcorn void populations.

In order to find a statistically-significant void catalog, we must find a suitable set of parameters.
Choosing values for $\Delta_v$ and $R_{\max}$ is generally straight-forward.
Due to the close connection between spherical void finders and \pvf, we tend to inherit $\Delta_v$ from the spherical theory, which predicts voids as extended structures with $\Delta_v \approx -0.8$ in all of their volume. 
$R_{\max}$ on the other hand can be derived from empirical evidence, using information such as the tracer type or the mean tracer density we can expect the voids to be smaller than a certain upper-bound.
Our choice for this experiment was $R_{\max}=45 \, h^{-1}\mathrm{Mpc}$, which ended up being far larger than the largest found sphere; but had it been the other way around, the effects would've still been traceable in the results.

Choosing an appropriate value for $R_{\min}$ is less straightforward since, at small scales, the spherical void finder (SVF) can produce void abundances that deviate from the expected physical distribution.
The void hierarchy theory predicts that the \vsf follows a Schechter-like form \citep{2004.Seth.Hierarchy,1976.Schechter}, describing the abundance of voids across the physically relevant range of scales.
In practice, however, this behavior is not necessarily recovered down to arbitrarily small radii.
These scales are predominated by the limited spatial information provided by the discrete tracer distribution, spherical void fragmentation and other finder-specific properties that collectively produce an artificial regime that we encompass under the term \textit{shot noise}.

To determine the transition between these regimes, we first select a deliberately conservative provisional minimum radius $\hat{R}_{\min}$ based on the mean inter-tracer separation,

\begin{equation}
\langle r \rangle = (\bar{\rho})^{-1/3} \, \text{.}
\end{equation}

In our simulation, $\langle r \rangle = 512/500 \approx 1 \, h^{-1}\mathrm{Mpc}$, which led us to adopt $\hat{R}_{\min}=2.5 \, h^{-1}\mathrm{Mpc}$; although values up to approximately three times the mean separation provide similarly safe initial choices in practice.
Then, we execute the spherical void finder using $\Delta_v$, $R_{\max}$, and $\hat{R}_{\min}$, allowing the resulting catalog to retain voids down to scales where shot-noise effects may become relevant.

The resulting \vsf exhibits two distinct regimes.
At sufficiently large radii, the measured abundance follows the expected Schechter behavior, whereas at smaller radii it departs systematically from this form; empirically found to follow a power-law in the current implementations \footnote{This behavior was different back when we used the more unconstrained random walks as the recentering strategy. We explain the benefits of the current method in \autoref{sec:concl}.}.
We therefore model the \vsf as a continuous piecewise function composed of a power-law at small radii and a Schechter function at large radii, split by the \textit{shot noise radius} $R_{\text{shot}}$:

\begin{equation}
    \frac{dn}{d \log R} = \begin{cases}
        b R^a & R < R_{\text{shot}} \\
        \phi^* \left(\frac{R}{R^*}\right)^{\alpha} e^{-R/R^*} & R \ge R_{\text{shot}} \text{.}
    \end{cases}
\end{equation}

\begin{figure}
    \includegraphics[width=1.0\columnwidth]{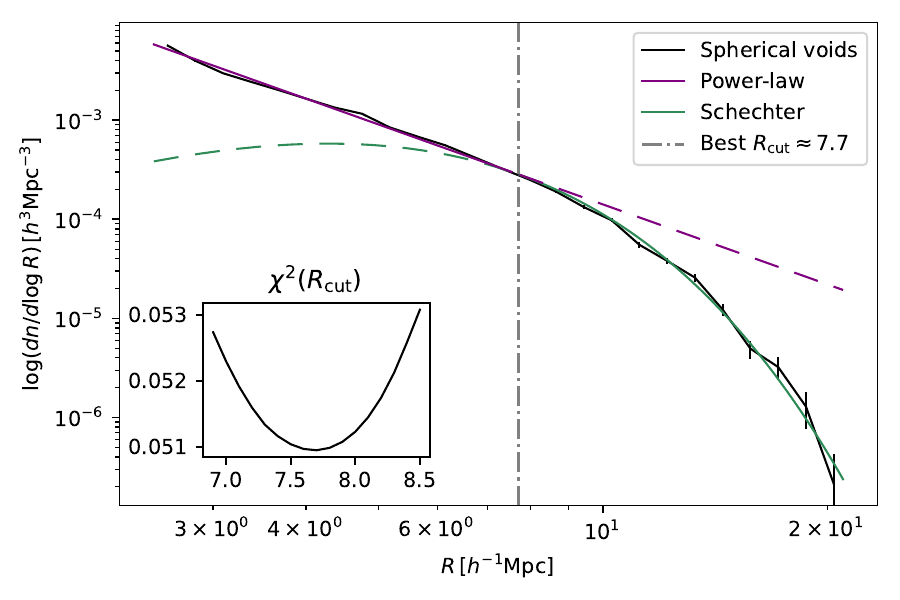}
    \centering
    \caption{Empirical \vsf for a set of spherical voids along with a piecewise fit of two curves divided by a cutoff point $R_{\text{cut}}$.
    The main frame shows the $R_{\text{cut}}$ that minimizes fitting loss.
    The frame in the bottom-left edge shows the loss of nearby candidate cutoffs.
    }
    \label{fig:rshot}
\end{figure}

In order to fit a continuous model, we adjusted these regimes to match at the point $\left( R_{\text{shot}},y_{\text{shot}} \right)$, resulting in a coupled re-parameterization of five dependent parameters: $R_{\text{shot}},y_{\text{shot}},a,\alpha$, and $R^*$.
Then, we select a set of candidate cutoff radii and fit the curve on the four remaining parameters.
We identify the shot-noise radius $R_{\mathrm{shot}}$ as the cutoff radius that minimizes $\chi^2$ fitting score, and adopt this value as the minimum radius to be used by the popcorn stage,

\begin{equation}
R_{\min}=R_{\mathrm{shot}}.
\end{equation}
The procedure is illustrated in \autoref{fig:rshot}, where we fitted a \vsf produced with $30$ logarithmic bins of the spherical void radii.

\begin{figure}
    \includegraphics[width=1.0\columnwidth]{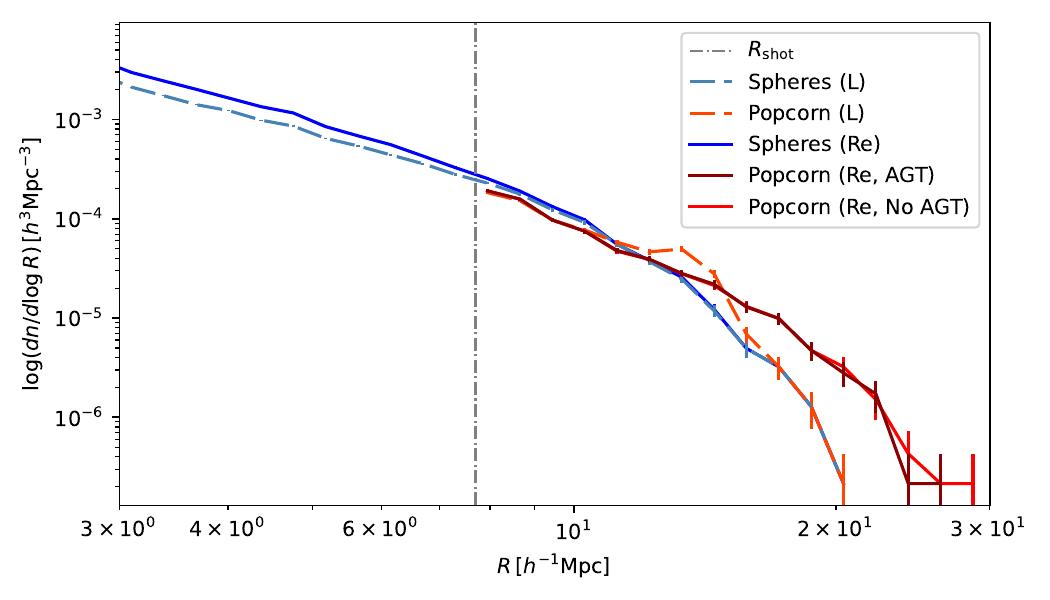}
    \centering
    \caption{Empirical VSFs for the different configuration of finders and implementations of \pvf.
    Blue-ish colors are used for spherical algorithms and red-ish for the popcorn variants.
    Dashed lines represent the legacy implementation (L) while solid lines refer to the rewritten implementation (Re).
    The ``AGT'' popcorn variant used a lossy ``adaptative growth termination'' criterion that strips raising voids early in exchange for performance.
    }
    \label{fig:vsfs}
\end{figure}

Once the appropriate parameters have been determined, we complete the pipeline by running the popcorn stages on the previously obtained spherical void catalogs.
To assess the quality of our reimplementation, we perform these runs with both codebases, together with an additional variant in which the adaptative growth termination criterion (discussed in \autoref{subseq:convergence}) is disabled.
All of these runs result in the \vsf curves shown in \autoref{fig:vsfs}, where the popcorn voids were binned using their \textit{effective radius}:

\begin{equation}
    R_{\text{eff}} = \left(\frac{3V}{4 \pi}\right)^{1/3} \text{.}
\end{equation}

On the right-side of $R_{\text{shot}}$, we can immediately see that the spherical void curves converge at around $R= 20 \, h^{-1} \mathrm{Mpc}$, while some of the popcorn curves reach just shy of $R = 30 \, h^{-1} \mathrm{Mpc}$.
This means that the popcorn algorithm was able to effectively expand fragmented or incomplete spherical underdensities into the extended void structures.
However, one of the popcorn curves fell behind, landing close with the spherical voids.
This occurred in versions following the initial release of the legacy implementation and was caused by its aggressive growth termination criterion.
Rather surprisingly, the early tests we performed using the legacy code with dark matter halos did not exhibit this close Sphere-Popcorn curve alignment we observed here with particles.
In contrast, the new adaptive termination criterion is able to closely replicate the ``original'' behavior past $R= 20 \, h^{-1} \mathrm{Mpc}$ until the last three bins, which in total accumulate to only four voids.

On the left-side of $R_{\text{shot}}$, we can see another divergence between the spherical void curves, which is caused by a flaw in its overlapping cleanup phase.
As we explained in \autoref{subseq:sph_overlap}, a ``clean'' spherical void configuration is sensitive to the ordering of overlap resolutions.
This property heavily constrains the reach of parallel solutions since they should carefully avoid processing overlapping chains at the same time.
Instead, the legacy implementation approached this in a two phase process: a stochastic preliminary filter, where large sets would be divided in chunks and processed in parallel if the maximum relative radius difference was below $5\%$; and a deterministic sequential pass that would process the rest of the chunk.
While this arbitrary threshold was designed to minimize any statistically noticeable effects, it nonetheless introduced a visible bias in this specific case.
It's no coincidence that this effect is more prominent on the lower ends of $R_{\text{shot}}$, as they represent the vast majority of voids; but we can't rely on this behavior being consistent on $R > R_{\text{shot}}$, which is why we re-implemented the cleanup procedure from scratch in the reimplementation.

\subsection{Performance}

\begin{figure}[!ht]
    \includegraphics[width=0.99\columnwidth]{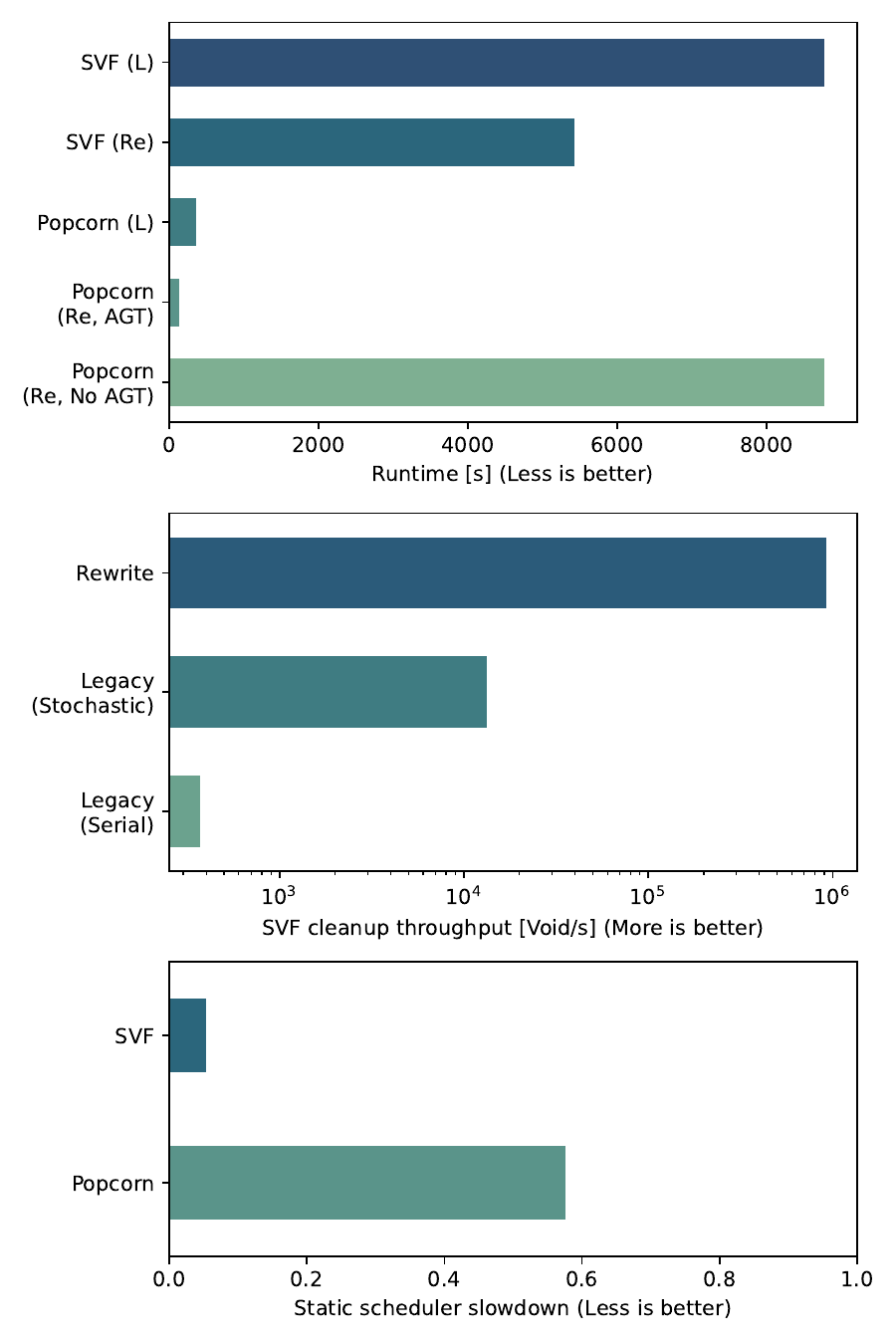}
    \centering
    \caption{\pvf performance analysis.
    The top panel shows the wall times of the legacy (L) and rewritten (Re) versions of SVF and Popcorn, with an additional Popcorn variation using the adaptive growth termination criterion (AGT).
    The mid panel shows the processing speed of the different overlapping spherical void cleaning strategies, where we subdivide the legacy strategies in a parallel stochastic version and a pure serial version.
    The bottom panel shows the slowdowns produced by the static scheduler, which enables deterministic execution.
    }
    \label{fig:perf}
\end{figure}

With a statistical assessment we successfully identified and mitigated various deficiencies in the legacy codebase.
To validate the broader impact of these interventions, we close with a rigorous performance comparison, demonstrating how these structural changes influenced execution speed.
The performance analysis evaluates three aspects of the rewritten implementation: overall pipeline runtime, the cost of spherical-void overlap removal, and the impact of deterministic scheduling.
The system used for these metrics was the \serafin cluster from the supercomputing unit of Universidad Nacional de Córdoba ``UNC Supercómputo'' \footnote{\url{https://supercomputo.unc.edu.ar/}}, featuring $2\times$ AMD EPYC 7532 CPUs per node and $256$GiB of RAM.
The code was compiled with \texttt{gcc} 12.3.0 and \texttt{openmpi} 4.1.6.
A complete summary of this analysis is shown in \autoref{fig:perf}.


\textbf{Pipeline runtime}:
We first compare the wall-clock runtime of the complete identification pipelines.
Since the legacy and rewritten implementations differ in their executable stages and post-processing, we combine the main identification and subsequent processing stages into a single runtime for each configuration.
For the Spherical Void Finder, the rewritten implementation reduces the runtime from $8767.8\text{ s}$ in the legacy version to $5420.9\text{ s}$, corresponding to a $1.62\times$ speedup.
This improvement provides a substantially faster input catalogue for the subsequent popcorn formation stage.

The popcorn results, however, illustrate the importance of considering the amount of work performed by each configuration.
The legacy implementation completes the pipeline in $360.8\text{ s}$, whereas the rewritten implementation without early growth termination (\texttt{Popcorn (Re, No AGT)}) requires $8773.4\text{ s}$.
This difference does not indicate a regression in the rewritten algorithm itself, but rather exposes the substantial amount of redundant growth work avoided by the legacy termination heuristic.
When the adaptive growth termination criterion is enabled, the rewritten implementation completes in only $139.4\text{ s}$, corresponding to a $62.9\times$ reduction relative to the unconstrained rewritten version.
This result also highlights the importance of the load distribution in the rewritten Popcorn implementation: in the unconstrained configuration, approximately $99.7\%$ of the runtime is spent processing only the final eight voids out of the $3408$ processed in the experiment.
The termination criterion therefore provides not only a reduction in the total amount of work, but also prevents a small number of pathological growths from dominating the execution time.

\textbf{Spherical void cleanup}:
Our next experiment was a performance evaluation of the spherical void overlap removal algorithms.
Because the implementations processed a different number of voids on each run, raw runtime is not a meaningful comparison.
Instead, we measured cleaning throughput, defined as the number of processed voids per second.

The legacy procedure that cleaned $6.3$ million voids in $480.5\text{ s}$, resulting in a throughput of $1.33\times10^4\text{ void/s}$.
Contrastingly, the rewritten implementation cleaned $6.8$ million voids in $7.3\text{ s}$, reaching a throughput of approximately $9.18\times10^5\text{ void/s}$.
Overall, this corresponded to approximately $69\times$ improvement.

For completeness, we also isolated the serial portion of the legacy implementation, excluding the timing of its stochastic pre-processing stage.
From the original $6.3$ million voids, only $180$ thousand remained, taking $458.9\text{ s}$ to process, resulting in a throughput of approximately $373\text{ void/s}$.
This illustrates the substantial overhead associated with the legacy overlap removal strategy, if we had just removed pre-processing to fix the bias from \autoref{fig:vsfs}, cleanup would have taken just shy of 5 hours, far outrunning the time it took to find those voids.
The rewritten procedure therefore provides a considerably more efficient implementation while simultaneously eliminating the stochastic behaviour introduced by the legacy pre-processing stage.

\textbf{Scheduling overhead}:
In this work we introduced a deterministic mode for the SVF, allowing for complete pipeline replicability.
This required the addition of a new static scheduling strategy, which ensures all parallel threads process input in the same order.
This final test evaluated the cost of this static scheduler over the original dynamic scheduler in void identification.
In the case of SVF, we only spot an approximate runtime increase of $5\%$ under static scheduling.

Although Popcorn itself is deterministic independently of the scheduling strategy and random seeds, it is included in this experiment to characterize the potential scheduling overhead in its growth workload.
In this case, the effect is considerably larger, with a runtime increase of approximately $58\%$.
These results indicate that static scheduling can introduce a measurable load-balancing cost, particularly for the highly irregular popcorn workload.
Nevertheless, there's negligible overhead in the case of SVF, which is the only step where this choice of scheduling is relevant.

\section{Summary and conclusions}
\label{sec:concl}

In this work, we revisited the implementation of the Popcorn Void Finder with the goal of improving its performance, maintainability, reproducibility, and statistical reliability.
Rather than treating the existing implementation as a black box, we reconstructed and documented the complete void-identification procedure, from the initial spherical candidates to the construction and filtering of popcorn voids.
This provided a detailed description of the algorithm and its underlying design choices, while also allowing us to identify and evaluate several sources of computational and statistical behavior in the original implementation.
The spherical and popcorn identification algorithms were revisited and reimplemented with a focus on separating their conceptual stages and reducing unnecessary complexity.
Beyond the algorithms themselves, the rewritten implementation improves the practical usability of \pvf.
HDF5 support was made optional, allowing configurations that do not require HDF5-based input formats to be compiled without this dependency.
Reproducibility was also introduced into the spherical finder through explicit random seeds and deterministic scheduling, allowing executions to be repeated under controlled conditions.

Static analysis of the codebase showed a reduction in total source-code size, per-function size, cyclomatic complexity, and duplicated code.
These changes provide empirical evidence that the rewritten implementation is not only more efficient in selected workloads, but also substantially easier to maintain and extend.
The statistical analysis showed that the rewritten implementation provides a more consistent basis for statistical analyses while preserving the ability of popcorn voids to represent non-spherical structures.
A performance evaluation revealed a $1.62\times$ speedup on the new spherical void finder, while the rewritten Popcorn implementation combined with a new adaptive growth termination strategy reached a $62.9\times$ speedup with a small loss in statistical quality.
These results emphasize that performance in Popcorn is strongly dependent on the amount and distribution of work performed, rather than solely on the cost of individual algorithmic operations.

This work also revealed a substantially different small-scale behavior in the Void Size Function.
The previous implementations \citep{2023.paz.popcorn, 2015.ruiz.cluesiii} employed a random-walk recentering procedure whose parameters, such as the step size and number of steps, directly affected the small-scale shape of the resulting \vsf.
Consequently, the location and strength of the transition into the shot-noise regime could vary with internal algorithmic parameters that were not directly exposed to the user, making the determination of the shot noise less robust.
The in-cell recentering strategy adopted in the present implementation produces a qualitatively different and more regular behavior.
The small-scale regime can be empirically described by a power law, followed by the expected Schechter form at larger radii, allowing the shot noise to be defined directly from the transition between these two regimes.
Importantly, this characterization does not introduce an implicit dependence on hidden parameters, providing a more reproducible basis for selecting the minimum reliable void scale.

Along with \pvf, we released a new open-source Python API \textit{pvfpy} \footnote{\url{https://gitlab.com/void-research-labs/pvfpy}} that is able to read and process output identification files.
In particular, void visualizations (\autoref{fig:pop_heuristics}), shot noise determination (\autoref{fig:rshot}), and Void Size Functions (\autoref{fig:vsfs}) were all produced with this API.
A recent feature introduces an alternative representation of popcorn voids as prolate ellipsoids, derived from their spatial distribution of member spheres.
This provides a compact description of their non-spherical geometry in terms of their principal dimensions and orientation, facilitating shape characterization and comparisons between voids.
The complete software stack is backward-compatible, allowing configuration and output files from the legacy \pvf to be used with the renewed codebase.
Overall, this work demonstrates the value of revisiting the original implementation beyond its initial scientific results.
By improving its maintainability and reproducibility while systematically evaluating its algorithmic choices, we establish a more reliable foundation for future developments and statistical applications of popcorn voids.

\section{CRediT authorship contribution statement}

\textbf{A. R. Schachner:} Writing – original draft, Software, Investigation, Formal analysis.
\textbf{D. J. Paz:} Writing – review and editing, Software, Supervision, Methodology, Validation.
\textbf{J. B. Cabral:} Writing – review and editing, Supervision, Methodology, Validation.
\textbf{J. A. Rico Gallego:} Supervision, Funding acquisition.
\textbf{C. M. Correa:} Writing – review and editing, Supervision.
\textbf{A. N. Ruiz:} Writing – review and editing, Software, Supervision.
\textbf{S. R. Gualpa:} Software.

\section{Declaration of generative AI and AI-assisted technologies in the manuscript preparation process}

During the preparation of this work, the author(s) used OpenAI--ChatGPT, Google--Gemini, and Anthropic--Claude in order to improve the clarity and correctness of the text.
After using this tool/service, the author(s) reviewed and edited the content as needed and take(s) full responsibility for the content of the published article.

\section{Declaration of competing interests}

Alvaro Roy Schachner reports financial support was provided by Sistema Extremeño de Ciencia, Tecnología e Innovación (SECTI). If there are other authors, they declare that they have no known competing financial interests or personal relationships that could have appeared to influence the work reported in this paper.

\section{Acknowledgments}
\label{sec:awk}

This work was supported by Consejo Nacional de Investigaciones Científicas y Técnicas (CONICET) and co-funded by the European Union through the European Regional Development Fund (ERDF) via the ``Consejería de Educación, Ciencia y Formación Profesional de la Junta de Extremadura'' (Ref. GR24208).
This work used computational resources from UNC Supercómputo (CCAD) \footnote{\url{https://supercomputo.unc.edu.ar}}, which is part of SNCAD, Argentina.

\section{Code availability}

The \pvfl source code is publicly available at \url{https://gitlab.com/void-research-labs/popcorn-void-finder} under the BSD 3-Clause License.

\section{Data availability}

The data supporting the statistical results presented in this work have been deposited in Mendeley Data and are available to reviewers through the private dataset link provided during submission.
The dataset will be made publicly available upon publication of the associated article.





\bibliographystyle{elsarticle-num}
\bibliography{refs}







\end{document}